\documentclass[12pt,preprint]{aastex}

\usepackage{graphicx}
\usepackage{dcolumn}
\usepackage{bm}

\usepackage{amssymb,epsfig}

\begin{document}

\title{A compact star rotating at 1122 Hz and the r-mode instability} 

\author{
Alessandro Drago\altaffilmark{1}, 
Giuseppe Pagliara\altaffilmark{2} and 
Irene Parenti\altaffilmark{1}}

\altaffiltext{1}{Dipartimento di Fisica, Universit{\`a}
di Ferrara and INFN, Sezione di Ferrara, 44100 Ferrara, Italy}
\altaffiltext{2}{Inst. Theoretische Physik, Goethe Universit\"{a}t,
D-60438, Frankfurt am Main, Germany and INFN, Italy}

\begin{abstract}
We show that r-mode instabilities severely constraint the composition
of a compact star rotating at a sub-millisecond period. In particular,
the only viable astrophysical scenario for such an object, present inside
the Low Mass X-ray Binary associated with the x-ray transient XTE J1739-285, 
is that it has a strangeness content. Since previous analysis indicate that 
hyperonic stars or stars containing a kaon condensate are not good candidates, 
the only remaining possibility is that such an object is either a strange 
quark star or a hybrid quark-hadron star. We also discuss under which conditions 
sub-millisecond pulsars are rare.
\end{abstract}
 
\keywords{star: neutron --- star: rotation --- r-mode --- quark star --- 
submillisecond pulsar --- gravitational waves}

Very recently \citet{Kaaret:2006gr} reported the evidence of a
X-ray transient with a pulsed component of the emission having a
frequency $f=1122 \pm 0.3$ Hz. This signal is interpreted as due to the
rotation of the central neutron star. As such this object would be the
most rapidly rotating compact star discovered up to now. 
This single observation clearly needs to be confirmed, maybe by the
analysis of future X-ray transients of the same object.
The implications of this rapid rotation on the Equation of State (EOS) and in
particular on the allowed values of the mass and the radius have been
discussed in \citet{Lavagetto:2006ew} and in \citet{Bejger:2006hn}. 
Here we discuss the problem of the stability respect to r-modes of such 
a rapidly rotating object (for a review on r-modes see \citet{Andersson:2000mf}). 

It has been widely discussed in the literature the possibility
that r-mode instabilities can very efficiently drag angular momentum
from a rotating compact star if its temporal evolution in the
$\Omega-T$ plane (angular velocity and temperature) enters the r-mode
instability window, see e.g. \citet{Andersson:1998xt} and \citet{Friedman:1998uh}. 
Therefore huge regions of the $\Omega-T$ plane are excluded. Moreover, the size 
and position of that window is strictly related to the composition of the star,
since it is strongly dependent on the value of bulk and shear viscosity. 
It is particularly important to recall that for stars containing strangeness, 
as hyperon stars \citep{Lindblom:2001hd}, hybrid stars \citep{Drago:2003wg}
and strange quark stars \citep{Madsen:1999ci} there is also a contribution 
to the bulk viscosity associated with the formation of strangeness. Due to this, 
the instability window splits into two parts: one which starts at temperatures 
larger than $(7 \pm 3)\, 10^{9}$~K (High Temperature Instability Window 
HTIW) and a lower temperature window at temperatures smaller than 
$(5 \pm 4) \, 10^{8}$~K (Low Temperature Instability Window, LTIW)
\footnote{Here and in the following we are discussing the temperature of the 
region of the star where r-mode instabilities can develop. Typically it is 
located at a depth of a few kilometers below the surface.}. 
Concerning the left border of the LTIW, its position is regulated by the 
shear viscosity and by the so called viscous boundary layers located at 
the interface between the crust and the fluid composing the inner part of the star 
\citep{Bildsten:1999zn}. In particular, in bare quark stars, either composed 
by normal or by superconducting quark matter, due to the absence of a 
significant crust, the left border of the LTIW extends to much lower 
temperatures and the LTIW has a minimum corresponding to a significantly 
lower temperature than in the case of stars containing a crust. 

In this Letter we discuss in which region of the $\Omega-T$ plane the compact 
star originating XTE J1739-285 is most likely located, due to its composition. 
Let us start by discussing the simplest possibility, i.e. that the object is 
a neutron star. In principle a neutron star can rapidly rotate in two cases, 
either if its temperature is very large, above a few MeV, or if it is recycling, 
spinning up due to mass accretion (see Fig.~1).
Concerning the first possibility, a hot neutron star would be a newly born 
one, since the time needed to cool below one MeV is of the order of one minute.
This is clearly not the case of the stellar object under discussion. 
Concerning recycling, it should take place on the left side of the instability 
window, located at lower temperatures (see Fig.~1). 
The main result of the analysis of \citet{Andersson:2000pt}, revisiting previous 
analysis \citep{Levin:1998wa,Bildsten:1999zn} is that a neutron star can never 
spin-up to a rotational period shorter than $\sim$ 1.5 ms. This result is based 
on the estimates of the temperature and of the mass accretion rates of Low Mass 
X-ray Binaries \citep{Brown:1997ji,Bhattacharyya:2001sw} indicating that 
temperatures lower than $\sim 10^8$~K cannot be reached. 
Therefore a sub-millisecond neutron star cannot be present at the center of a 
Low Mass X-ray Binary (LMXB).
It is important to remark that this conclusion is confirmed by more recent 
analysis, taking into account the composition stratification of the rigid crust 
\citep{Glampedakis:2006mn} and discussing the nonlinear development of the 
r-mode instability \citep{Bondarescu:2007jw}.

Let us now discuss the case in which the compact star contains strangeness. 
It has been shown, in the case of hyperon stars 
\citep{Wagoner:2002vr,Reisenegger:2003cq}, and for hybrid stars 
\citep{Drago:2004nx} (strange quark stars also have a similar behavior), that 
due to the large value of the bulk viscosity associated with the non-leptonic 
weak decays, there are two windows of instability, the LTIW and the HTIW 
introduced above.
The HTIW does not affect significantly the angular velocity of the star because 
the cooling of a newly born star is so fast that there is not enough time for 
the r-mode instability to drag a significant fraction of the angular momentum.
Therefore, the star exits the HTIW with an angular velocity close to the initial 
one. When the temperature drops down to a few $10^8$ K the star reaches the LTIW 
and it starts to lose angular momentum due to r-mode instability. In Fig.~1 
we show examples of the instability windows in the case of a pure quark star 
and of a hybrid star. Notice that the position of the LTIW depends rather
strongly on the mass of the strange quark $m_s$. For large values of 
$m_s$ the instability windows shrink considerably.  
In our analysis we have considered two possibilities concerning the value of $m_s$:
a small value $\sim$ 100 MeV, of the order of the strange quark current mass,
and a large value $\sim$ 300 MeV, similar to what has been obtained in 
NJL-like models of quark matter in which a density-dependent constituent
mass has been computed \citep{buballa}.

Concerning the left side of the LTIW, 
it is easy to see that it cannot be used to accommodate a submillisecond pulsar.
Indeed, 
in the case of hybrid or hyperonic stars the left side of the LTIW is similar 
to the one of neutron stars, discussed above,
and we can apply here the 
same analysis done for a neutron star and concerning the minimal temperature of the 
compact object at the center of a LMXB. Therefore also hybrid or hyperonic stars 
cannot recycle to frequencies significantly exceeding 700--800 Hz if they are
to the left of the LTIW.
Moreover, in the case of quark stars the absence of 
viscous boundary layers implies that they can
rotate rapidly only on the right side of the LTIW (a very 
light crust suspended over the surface of a quark star is irrelevant from the 
viewpoint of r-mode instability, although it can be important for the production 
of x-ray bursts, see \citet{Page:2005ky}).
In conclusion, stars containing strangeness can rotate at
submillisecond periods only if they are to the right of the LTIW.

Let us now discuss the results of our analysis also considering the constraints 
posed onto the EOS by the mass shedding limit. The main result of 
\citet{Lavagetto:2006ew} and of \citet{Bejger:2006hn} is that soft EOSs, and 
in particular the ones based on hyperonic matter or on matter in which kaon 
condensation takes place, are rather unlikely. 
In fact, not only the range of allowed masses is rather small, but moreover
the only configurations satisfying the stability constraint for mass shedding
turn out to be supra massive \citep{Bejger:2006hn}, making it very difficult
to use these stellar structures in a mass accretion scenario.
This result confirms previous discussions of other astrophysical objects ruling 
out soft EOSs \citep{Ozel:2006bv}. On the other hand, stars containing 
deconfined quark matter are not excluded if the quark EOS is stiff enough 
\citep{Alford:2006vz}. We can therefore conclude that the object under 
discussion is either a quark or a hybrid star.

We can now study the temporal evolutions of the angular velocity of the star $\Omega$, of its 
temperature $T$ and of the amplitude of the r-modes $\alpha$ which are calculated by 
solving the set of differential equations given in \citet{Andersson:2001ev} 
(Eqs. 15-23-24). The only technical difference in our calculation is the 
inclusion of the reheating associated with the dissipation of r-modes by bulk 
viscosity, as discussed in various papers 
\citep{Wagoner:2002vr,Reisenegger:2003cq,Drago:2004nx}. 
Due to the reheating, the time needed for the star to cool down along the right 
border of the LTIW is rather long.
In the equation regulating the thermal evolution \citep{Andersson:2001ev} 
we have added a term: 
\begin{equation}
\dot E_{bv}=2E/\tau_{bv} \nonumber
\end{equation}
where $E \sim 10^{51} \alpha^2 M_{1.4} R^2_{10} P^{-2}_{-3}$ erg is the energy 
of the r-modes (the mass, radius and rotation period of the star are here 
expressed in units of 1.4 $M_{\odot}$, 10 km and in millisecond, respectively) 
and $\tau_{bv}$ is the bulk viscosity. Due to the large value of the bulk 
viscosity for strange matter at temperatures of a few $10^8$ K, the related 
dissipation of r-modes strongly reheats the star. 
The trajectory in the $\Omega-T$ plane describing the time evolution of the 
star follows essentially the border of the LTIW and the star keeps rotating as 
a sub-millisecond pulsar for a very long time,
strongly dependent on the value of $m_s$ and ranging from 
$\sim 10^2$ years when $m_s=$ 100 MeV to $\sim 10^8$ years when $m_s=$ 300 MeV, 
as shown in Fig.~2. 
In Fig.~3 we show the temporal dependence of the r-mode amplitude. 
Notice that, in the absence of reheating the time needed to slow-down to 
frequencies below 1 kHz is at maximum of the order of $\sim 10^4$ years,
and can be much shorter than a year if a small value of $m_s$ is adopted. 
In the past this branch of the 
instability window was only discussed in connection with very young pulsars 
because reheating due to bulk viscosity was not taken into account.

Before discussing the possible astrophysical scenarios for a submillisecond
pulsar inside a LMXB, it is important to remark two constraints
that a realistic model should fulfil. First, a LMBX is an old object,
with a typical age of $10^8$--$10^9$ years which is the time needed by the
companion (having a low mass) to fill its Roche lobe and to start accreting
on the neutron star. Therefore we need to provide a mechanism allowing
the central object inside the LMXB to rotate rapidly while being so old.
Second, submillisecond pulsars are certainly rather rare, and in particular
there is no evidence of a uniform distribution of pulsar rotational frequencies
extending from a few hundred Hz up to more than a kHz. Therefore a realistic model
should also indicate why most of the compact stars will not be detected
rotating at a submillisecond period. As we will show, these two constraints
can be satisfied in two different astrophysical scenarios: a first scenario 
in which a old hybrid or quark star is accelerated up to frequencies exceeding one kHz by mass
accretion and a second scenario in which the quark or hybrid star
is born with a submillisecond period and it is now spinning down by r-mode
instability (see Fig.~4).

The possible realization of these scenarios depends on two main
ingredients: 
\begin{itemize}
\item
the value of $m_s$, which, as we have already shown, regulates
the magnitude and position of the LTIW;
\item
the cooling rate of the central region of the star which determines the inner
temperature $T$ of a star which is accreting material from the
companion \citep{haensel}. 
\end{itemize}

Concerning the cooling rate, if strange quark matter is present in a compact star,
direct URCA processes are possible and therefore the cooling is (generally)
fast. It turns out from \citet{haensel} that in this case the inner temperature is
$\sim 5 \times 10^7$ K for a mass accretion rate of $\dot{M}=10^{-10}
M_\odot/$year and that the temperature scales as $T \propto \dot{M}^{1/6}$.
Another possibility recently proposed is that,
due to the formation of diquark condensates, URCA processes are 
strongly suppressed and the cooling turns out to be slow \citep{Blaschke:1999qx}. 
At the same time, bulk viscosity can still be large enough to suppress r-mode instability 
\citep{Blaschke:last}. In the following we discuss both possibilities, either
of a fast or of a slow cooling.

\noindent
{\it Let us consider first the case in which $m_s=300$ MeV.}  
\\The first
scenario, i.e. the star span up by accretion, can indeed be realized
if $T\ge 10^8$~K (see lower panel of Fig.~1). 
Such a temperature can be reached via reheating due to mass accretion. 
In the case of fast cooling a rather large value of mass accretion
rate is needed, $\dot{M}\sim 10^{-8} M_\odot/$year, and this stringent request can 
explain why submillisecond pulsars are rare.
A model in which the cooling is slow is instead excluded, because it would
be extremely easy to re-accelerate the star to very large frequencies
and therefore submillisecond pulsars would not be rare.
\\Also the second scenario in which the star is spinning down due to 
r-mode instabilities is possible. 
As shown 
in Fig.~2, the time spent by the star above
$1122$ Hz is of order of $10^8$ years and it is therefore compatible with the 
typical age of LMXBs. In this case submillisecond pulsars 
are rare because only a (small) fraction of newly
born compact stars can rotate with submillisecond periods (see e.g. the discussions in
\citet{Andersson:2000mf} and in \citet{perna}).
In this scenario the distribution of millisecond and submillisecond pulsars
is determined by the mass of the star at birth: if the mass is large enough,
strange quark matter can form in the core of the star, r-modes are efficiently damped
by the large bulk viscosity 
and the star can be a submillisecond pulsar at birth; if the mass of the star
is small, strange quark matter can not form, the bulk viscosity
of nucleonic matter can not damp r-modes and therefore in a short period
the star loses almost all its angular momentum \citep{Lindblom:prl}.

\noindent
{\it We consider now the case in which $m_s=100$ MeV.}   
\\The first scenario in which the star is spinning up by
accretion can be realized but $T$ must be at least $\sim 5 \times
10^8$ K for the star to be accelerated up to $1122$ 
(see upper panel of Fig.~4).  
Such a temperature can be reached but it needs strong reheating
due to a large mass accretion rate 
and slow cooling. 
\\The second scenario, i.e. the star is spinning down by r-mode instability, is
ruled out since, as shown in Fig.~2, the time spent by the star above
$1122$ Hz is of order of $100$ years and this is incompatible with the
typical age of a LMXB.

In conclusion, the two astrophysical scenarios proposed 
to explain a submillisecond pulsar inside a LMXB can both be realized by a quark or a hybrid
star if $m_s \sim 300$ MeV. Instead, if the $m_s$ is
small, only the first scenario is possible and it requires a slow cooling.
Our results are summarized in Table 1.

The main uncertainties in our analysis are due to the possible
existence of other damping mechanisms, taking place on the left side
of the LTIW. For instance magnetic fields can be important to suppress
r-mode instabilities \citep{Rezzolla:1999he}, but their effect is
probably negligible for frequencies exceeding $\sim 0.35\ \Omega_K$,
if the internal magnetic field is not larger than $\sim 10^{16}$ G.  
Obviously, even
larger uncertainties exist concerning quark matter.  As discussed
above, the bulk viscosity of quark matter strongly depends on the
strange quark mass and on the possible formation of a diquark
condensate \citep{Alford:2006gy}. Clearly, our analysis can provide
much needed constraints on the EOS of quark matter.

Finally, let us stress that the outcome of our analysis is that 
a compact star rotating at a submillisecond period
inside a LMXB
can only be a quark or a hybrid star. Future observations will be important to
clarify if
the object at the center of XTE J1739-285 constitutes indeed an example.

\acknowledgments
It is a pleasure to thanks Carmine Cuofano for many useful discussions.

\clearpage

\begin{table}
\caption{Microphysics and astrophysical scenarios}
\begin{tabular}{l|l|l|l|}
\hline
m$_s$ [MeV] & cooling & scenarios (I or II) & conclusions \\
\hline\hline
    &      &                                          &\\
300 & fast & I: O.K., $\dot M\ge 10^{-8} M_\odot / y$ & O.K. for \\
    &      & II: O.K., needs rapid rotation at birth & both scenarios \\
\hline
    &      &                                          &\\
300 & slow & I: no, submillisecond pulsars too frequent & excluded \\
    &      & II: O.K.                               & due to scenario I \\
\hline
    &      &                                          &\\
100 & fast & I: no, T is too low & not possible \\
    &      & II: no, slow down too  rapid & in both scenarios\\
\hline
    &      &                                          &\\
100 & slow & I: O.K., needs large $\dot M$ & O.K. for \\
    &      & II: no, slow down too rapid   & scenario I\\
\hline
\end{tabular}
\caption{Connection between microphysics (mass of the strange quark and cooling rate)
and astrophysical scenarios for
a submillisecond pulsar inside a LMXB.
The first scenario corresponds to a compact star which is spinning up due to
mass accretion; the second scenario to a compact star born as a submillisecond pulsar
and still in the process of slowing down.}
\end{table}

 \begin{figure}
    \begin{centering}
\epsfig{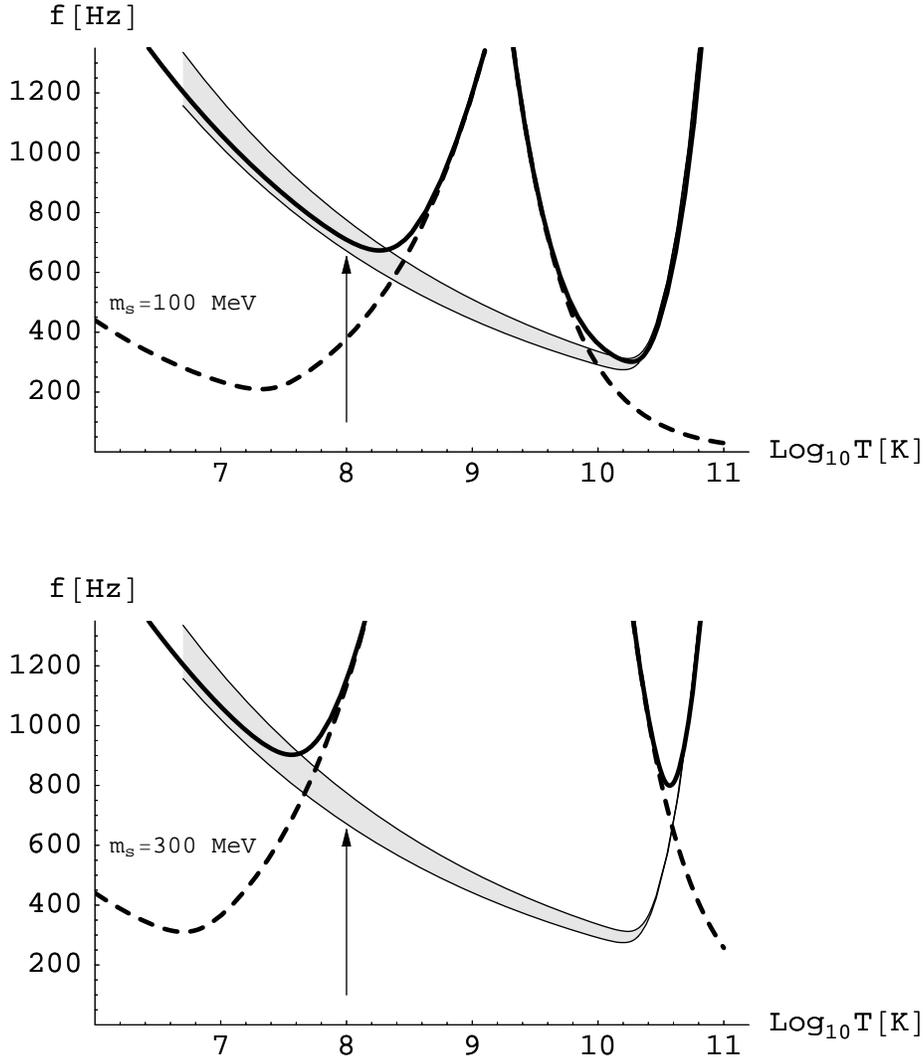}
    \caption{r-mode instability windows. The star is unstable and loses
angular momentum by emitting gravitational waves in the regions
above the instability lines displayed in figure. 
Thin solid lines correspond to neutron stars, for two extreme values 
of masses and radii allowed by the analysis done by \citet{Bejger:2006hn}. 
The shaded area in between is representative of intermediate values
of masses and radii. 
Dashed lines delimit the LTIW and the HTIW for strange quark stars. 
Thick solid lines correspond to hybrid stars. In both cases the 
mass and radius of the star are $M=2 M_{\odot}$ and $R=13$ km.
In the upper panel $m_s=100$ MeV and in the lower panel
$m_s=300$ MeV.
The bulk and shear viscosities are in general from \citet{Andersson:2000mf}.
Bulk viscosity of strange quark matter at high temperatures is from 
\citet{Madsen:1999ci}.
The arrow indicates the minimal temperature considered possible in LMXBs. 
The dot indicates the actual position of the compact star associated with 
XTE J1739-285, as it results from our analysis.
\label{instability}}
   \end{centering}
   \end{figure}

 \begin{figure}
    \begin{centering}
\epsfig{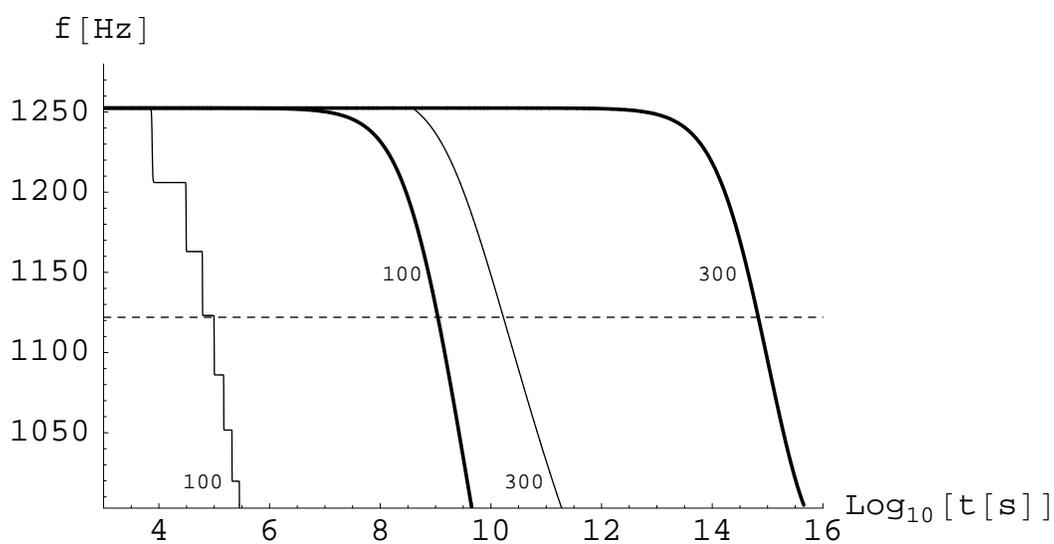}
    \caption{Time dependence of the rotational frequency of the compact star.
The thin lines take into account only the reheating due to shear viscosity, 
the thick lines take into account also the effect of reheating due to bulk 
viscosity. We show results for two different values of the strange quark mass, 
$m_s=$100 MeV and $m_s=$300 MeV. The dashed horizontal line
corresponds to a frequency of $1122$ Hz.
\label{freq}}
   \end{centering}
   \end{figure}

\begin{figure}[!ht]
\parbox{14cm}{
\centering
\scalebox{0.5}{
\includegraphics*[33,237][545,759]{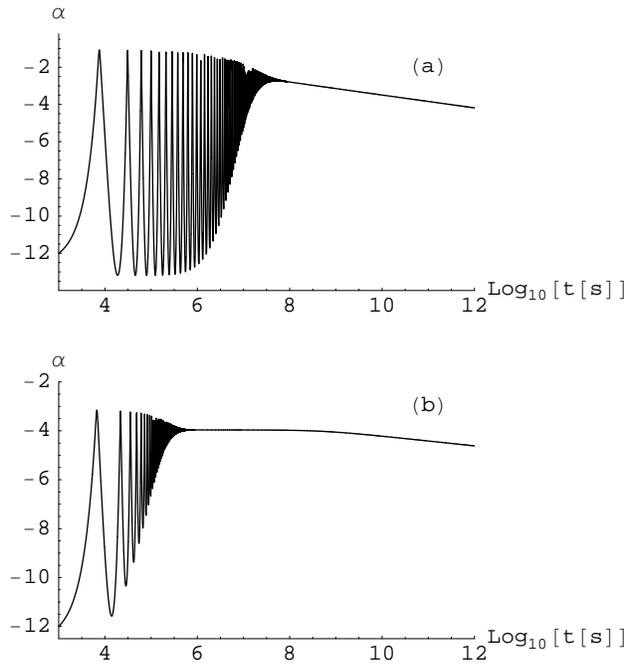}}
\caption{
Temporal evolution of the amplitude $\alpha$ of the r-modes. 
Results without reheating in panel (a) and taking into account reheating in 
panel (b). 
The amplitude of the r-mode is significantly damped when reheating is taken 
into account, the maximum amplitude is reduced by roughly 2 orders of magnitude. 
Therefore non linear effects should be suppressed.
\label{fig-alpha}}
}
\end{figure}

 \begin{figure}
    \begin{centering}
\epsfig{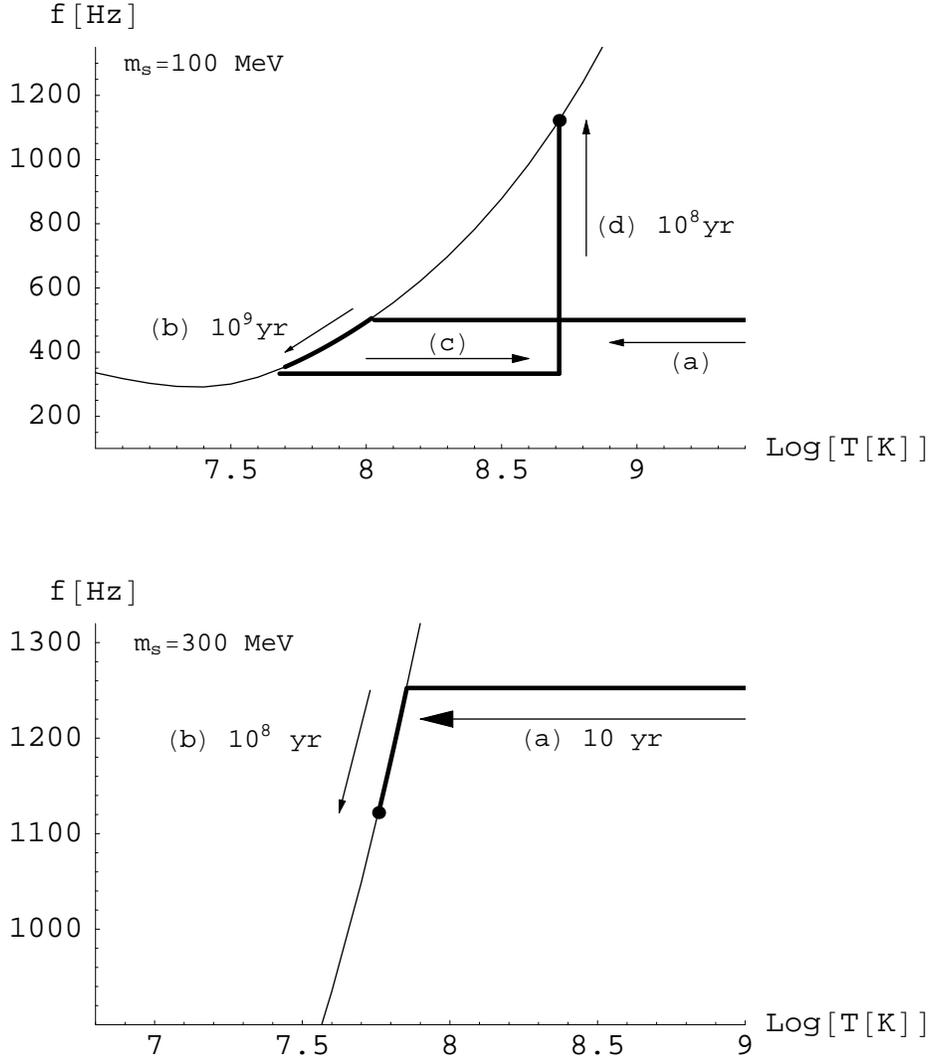}
    \caption{Upper panel: first scenario (see text). The star is born with a low
    angular velocity. After a cooling period at a constant
    angular velocity (line (a)), the star reaches the instability window and
    loses angular momentum during a long period, $\sim 10^9$ years (line (b)). 
After this long period
    the mass accretion process starts and the star is
    heated (line (c)) and re-accelerated (line (d)) to a frequency of 1122 Hz
    (indicated by the thick dot).
    Lower panel: second scenario (see text). The star is born with
    a large angular velocity. After a short period
    of cooling at a constant frequency (line (a)) the star enters the instability window and
    loses angular momentum via r-mode (line (b)). The frequency of the
    star can be larger than $1122$ Hz (indicated by the thick dot) for a period of $\sim
    10^8$ years. The mass and radius of the star are $M=1.4 M_{\odot}$ and $R=10$ km,
respectively.
\label{scenari}}
   \end{centering}
   \end{figure}

\end{document}